\documentclass[10pt,dvips]{article}
\usepackage{a4p}
\usepackage{multicol}
\usepackage{cite}
\usepackage{graphicx}
\usepackage{rotating}
\usepackage{epsfig}
\usepackage{picinpar}
\newcommand{\degrees}{^{\circ}}

\renewcommand{\baselinestretch}{0.99}
\newlength{\glositemwidth}
\newlength{\glositemindent}
{%
\small%
\begin{list}%
{~}%
{\setlength{\topsep}{0.0ex}%
 \setlength{\partopsep}{0.0ex}%
   \setlength{\itemsep}{-\parsep}%
 \addtolength{\itemsep}{0.35em}%
   \setlength{\leftmargin}{\glositemwidth}
 \addtolength{\leftmargin}{\glositemindent}  
   \setlength{\labelsep}{0.0ex}
 \addtolength{\labelsep}{\glositemindent}  
 \setlength{\rightmargin}{0ex}%
 \setlength{\labelwidth}{\glositemwidth}%
}%
}%
{%
\end{list}%
}%

\providecommand{\GLOSITEM}[2]%
{\item[{\makebox[\glositemwidth][l]{{\bf{#1}}}}]{#2\smallbreak}}

\newenvironment{myitemize}%
{%
\begin{list}%
{$\bullet$}%
{%
\setlength{\topsep}{0.0ex}%
\setlength{\itemsep}{-\parsep}\addtolength{\itemsep}{0.25em}%
\setlength{\partopsep}{0.25em}%
\setlength{\labelwidth}{2.5ex}%
\setlength{\labelsep}{1.0ex}%
\setlength{\leftmargin}{3.5ex}%
\setlength{\rightmargin}{0.0ex}%
}%
}%
{%
\end{list}%
}%

\date{\today}
\begin{document}

\vspace*{-4mm}

\begin{center}
{\LARGE\bf{On the use of Satellite Television \vspace*{4mm}\\
                in High Energy Physics}}
\end{center}

\vspace*{10mm}

\begin{center}
{\Large\bf{Lucas Taylor$^{1)}$~and~David O. Williams$^{2)}$}}
\end{center}

\vspace*{4mm}

{\small{%
\begin{center}
{\parbox{0.51\textwidth}{%
$1)$~~Northeastern University, Boston~~{\tt{(Lucas.Taylor@cern.ch)}}         \\[0.7mm]
$2)$~~CERN, Geneva~~{\tt{(David.O.Williams@cern.ch)}}     \\
}}
\end{center}
}}

\vspace*{10mm}

\begin{center}
{\Large{Invited talk at CHEP'98, Chicago, USA \vspace*{2mm}\\
             August 31 - September 4, 1998}}
\end{center}

\vspace*{15mm}

\begin{center}
{\bf\large{Abstract}}\\[4mm]
\parbox{0.715\textwidth}{%
\noindent%
This paper assesses the feasibility of exploiting commercial satellite television 
technologies to broadcast video signals and data from major 
High Energy Physics facilities to collaborating institutes throughout the world.}
\end{center}

\vspace*{8mm}

%
\section{Introduction\label{intro}}
%
%

\begin{figure}[!htbp]
\begin{center}
    \mbox{\epsfig{file=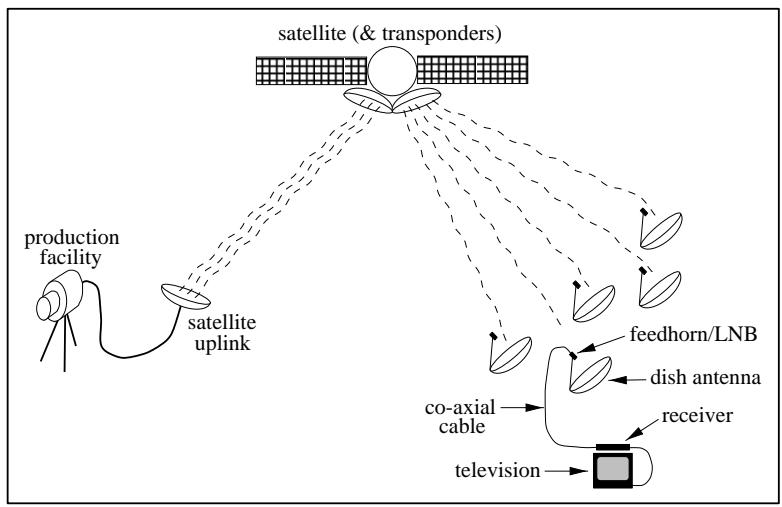,width=0.6\textwidth,clip=}}
  \caption{Schematic layout of a Satellite Television 
           broadcasting system.\label{fig:satellite}}
\end{center}
\end{figure}

Satellite Television (STV) broadcasting consists of the transmission of a 
video (or other) signal from an earth station to a geostationary satellite 
which subsequently broadcasts the signal back to the earth where it can be 
received over a wide area~\cite{ELBERT97A,PASCALL97A}.
The schematic layout of an STV system is shown in figure~\ref{fig:satellite}.
First the production facility provides a video signal to the 
satellite uplink transmitter.
A transponder on the satellite receives the uplink signal, modifies
the frequency, amplifies it, and rebroadcasts a polarised 
microwave signal back to earth. 
This signal is received using a parabolic antenna which relays it to an
analogue and/or digital receiver which converts the signal to a format suitable 
for display on a television set or recording on a sequential or random 
access medium such as video tape or compact disk. 
                
Satellite Television (STV) systems are used not only by television broadcasters 
but also by an increasing number of commercial and governmental organisations for
a multitude of purposes, including:
news feeds;
organisation-wide meetings and announcements;
training, referred to as Interactive Distance Learning (IDL);
and for liaison with the press, for example for the 
announcement of new products.
STV is especially attractive in that it provides a real-time video-bandwidth 
transmission mechanism which can reach arbitrarily large numbers of sites 
worldwide, at a cost which depends only weakly on the number of recipients.

The size, geographical distribution, and communications needs of the HEP 
community are not too dissimilar to those of large multi-national companies
and organisations which already make use of STV systems.
In recent years video-conferencing has become a widely-used tool in HEP. 
We anticipate a steady increase in the use of such IP-based Collaborative 
Working Tools as physicists come to rely more on such technologies and as 
more sophisticated commercial products become available~\cite{BUNN_CHEP97}.
In addition to such applications which operate over the fibre-dominated 
IP networks, there are a number of HEP activities which could profit from the
point-to-multipoint nature of STV broadcasting, including
seminars, major conferences, collaboration meetings, and training (IDL).

This study, which was carried out in the context of the ICFA 
Networking Task Force (NTF)~\cite{ICFA_NTF_REPORT_JULY98,WILLIAMS_CHEP98},
surveys STV technologies, assesses the potential of exploiting them in HEP,
and estimates the cost of a global STV system dedicated to HEP. 
Although HEP does makes use of point-to-point satellite links, 
for example between DESY and Moscow State University,
we do not consider such applications here.

%
\section{Satellite Technologies\label{technology}}
%
%
\subsection{Satellite Systems}

There are three classes of satellites according to the nature of their orbit.
``GEO'' satellites are in geostationary orbit above the equator with an altitude of 
approximately 36\,000\,km.
``LEO'' (Low Earth Orbit) and ``MEO'' (Medium Earth Orbit) satellites
are in non-geostationary orbits of up to 2\,000\,km and 
approximately 10\,000\,km respectively.
A number of multi-billion dollar LEO and MEO projects are already underway 
or planned for the start of the next century, involving systems of 
10's--100's of satellites.
These include 
IRIDIUM,   
GLOBALSTAR, 
ICO,       
ODYSSEY, and   
TELEDESIC\cite{TPN_MAY97_SHORT,WWW_SATMAGAZINE}.  
The lower orbits of LEO/MEO systems means there is less latency (delay) and more directed 
signals than GEO systems but that many more satellites are needed in order to 
provide global geographical coverage.
These systems are targeted at the personal communications market and offer, in conjunction
with integrated ground systems, such services as mobile telephony, fax, messaging, 
GPS, and broadband multimedia communications at costs of a few US\$/minute. 
Irrespective of the higher cost for transmissions, LEO/GEO satellites are inappropriate 
for STV due to their low orbit (and hence poor coverage) and the point-to-point 
(rather than broadcast) nature of their transmissions.
Therefore we do not consider them further.

In contrast to LEO/MEO systems, only three GEO satellites are required for 
global coverage, making them ideal for wide-area broadcasting of television signals.
As of February 1998 there were a total of 170 commercial GEO satellites 
with a further 155 foreseen by the end of the century, 
not including proposed higher frequency Ka-band systems~\cite{ALEWINE_SHORT}.
A GEO satellite can transmit multiple signals continuously, except for 
$O(10\,{\mathrm{minutes}}-1\,{\mathrm{hour}})$ per day
when the sun causes excessive noise. 
In terms of the geographical coverage by GEO satellite, the world may be conveniently 
divided into three main regions of longitude~\cite{WWW_SATCODX}:
Europe, parts of the FSU, Africa and the Middle East  ($69\degrees$E to $38\degrees$W);
North and South America ($41\degrees$W  to $139\degrees$W); and
Asia, parts of the FSU, and the South Pacific ($174\degrees$W to $74\degrees$E).
Satellites on the peripheries of these regions may, however, 
also beam signals simultaneously to two regions.
Pairs of GEO satellites transmitting on the same frequency are separated by 
approximately $2.5\degrees$ to avoid interference.
Often several satellites are co-located for the convenient 
reception of many channels with a static dish.
For example, in Europe a number of ASTRA~\cite{WWW_ASTRA} satellites are 
co-located at $19.2\degrees$E and a number of the Eutelsat~\cite{WWW_EUTELSAT}
 ``Hot Bird'' satellites are co-located at $13.0\degrees$E.

Satellite operations are typically managed by consortia of government and 
commercial enterprises or by large corporations.
Major satellite operators in Europe include: 
{{ASTRA}}~\cite{WWW_ASTRA}, operated by the 
{{Soci{\'{e}}t{\'{e}} Europ{\'{e}}enne des Satellites (SES)}};
{{Eutelsat}}~\cite{WWW_EUTELSAT}; and
{{Intelsat}}~\cite{WWW_INTELSAT}. 
Major satellite operators in the Americas include: 
{{Hughes}}~\cite{WWW_HUGHES}; 
{{PanAmSat}}~\cite{WWW_PANAMSAT}; 
{{Loral}}~\cite{WWW_LORAL}; 
{{Orion}}~\cite{WWW_ORION}; 
{{Lockheed Martin}}~\cite{WWW_LOCKHEED} 
(in collaboration with {{Intersputnik}}~\cite{WWW_INTERSPUTNIK}); 
and {{COMSAT}}~\cite{WWW_COMSAT} (recently acquired by Lockheed Martin). 
In addition to such primary operators, there are ``satellite service providers''
who sell bandwidth for arbitrary amounts of time and 
offer services such as business television, video conferencing, 
audio conferencing, and enhanced FAX services with multiple recipients, 
using combinations of satellite and fibre-optic channels. 
One such example, used below in our cost estimates, is 
{{Global Access}}~\cite{GLOBALACCESS_COSTS} which is part of 
the {{Williams Communications Group}}~\cite{WWW_WILLIAMS}.  

\subsection{Satellite Broadcasting}
The signal originating on earth is first transmitted to the satellite from 
an uplink station, which may be either a permanent dish antenna or a mobile unit 
typically consisting of a van with a dish on the roof.
The signal is received by one particular transponder on the satellite which 
alters the frequency of the signal, amplifies it, and rebroadcasts a  
polarised signal back to earth. 
The microwave frequencies used for satellite transmissions are subdivided 
into bands~\cite{ELBERT97A}.
The transmission of STV signals in Europe and the Americas is in the 
Ku-band with a frequency of $10-13$\,GHz.
To effectively double the available bandwidth, two orthogonal polarisations 
are generally used for the same frequency band (either horizontal and vertical planar
polarisation or left and right circular polarisation). 

The power for each transponder (typically 50\,W) depends on the total number 
of transponders (typically 24--48 per satellite) and the total power generated 
by the solar panels of the satellite (typically several kW).
The transponder may direct the signal at a specific location or
spread the beam over a broad geographical area.
The maximum size of the transmission region, or footprint, is constrained by 
the power of the transponder and the sensitivity (area) of the target dish 
antennae which receive the signal.
The strength of the signal is expressed in terms of 
Effective Isotropic Radiated Power (EIRP).
Since transmissions are not isotropic, satellite operators publish the 
footprints of their transmissions which consist of 
EIRP values superimposed on a map.
These values are typically expressed in terms of dbW (decibel-Watts) which
denote the power of the signal at a given location, relative to one Watt. 
Footprints may quote nominal transponder power whereas the actual power available 
for a given channel may be lower, for example if several channels share the 
same transponder.
The ``power back-off'' of the transponder in decibels (dB), which is also 
available from the satellite operator, should in general 
be subtracted from the nominal EIRP when estimating effective power.
Also, one must ensure that the reference footprint applies to the appropriate 
beam mode of the transponder (e.g. either ``Widebeam'' or ``Superbeam'' 
for Eutelsat transmissions).

The signals transmitted may be either analogue or digital.
The latter is becoming preferred since the use of compression 
techniques, such as MPEG~\cite{ELBERT97A,PASCALL97A}, reduces the 
bandwidth requirements by 1-2 orders of magnitude.
Digital technology permits not just video but arbitrary data 
to be broadcast.
The problems of higher bit error rates and latency of satellite links compared to 
fibre may be tackled using Forward Error Correction (FEC) systems  such as that 
based on the Reed Solomon algorithm~\cite{PASCALL97A} and/or a 
suitably adjusted TCP/IP implementation~\cite{WWW_SEPMEIER_97A}.

\subsection{Reception} 

STV transmissions may reach the viewer via a number of routes.
A terrestrial station may receive and retransmit the signal either 
``on the air'' or via cable.
Alternatively it may be received on a satellite antenna dish at the viewers location.
Since the HEP community is relatively small and highly distributed 
we only consider the latter option.
Dish antenna reception systems may be of two sorts:
``Direct to Home'' (DTH) reception in which the dish antenna serves a 
single television set; or 
``Satellite Master Antenna Television'' (SMATV) reception in which a single 
dish serves amplified signals to many individual users in a 
building or set of buildings.
DTH reception is appropriate for viewers such as HEP groups in universities
while SMATV may prove more appropriate for sites with multiple viewing rooms,
such as the major laboratories.
The technologies involved in DTH and SMATV are very similar, apart from 
slight differences in the size of the dish, the number of LNB's (described below),
the cabelling, and SMATV's need for additional amplification. 
The components required for either DTH or SMATV reception are readily 
available commodity items.

The signal is typically received from the satellite using a 
parabolic metal reflector dish.
The diameter of the receiving satellite antenna dish required 
for acceptable reception depends on:
the power output of the satellite transponder,
the beam mode of the transponder (e.g. wide or focused)
and the geographical location of the satellite dish within the footprint.
Acceptable reception is normally defined as a picture free of speckles
for 99.9\% of an average year.
This depends somewhat on the weather conditions, particularly in the tropics 
which suffer from signal losses due to rain.
In general, a larger dish means better reception quality.
From the EIRP one may determine the minimum size of dish required for 
acceptable reception using a conversion table provided by the satellite company.

A Low Noise Block (LNB) sits on the back of the feedhorn which 
receives the signal reflected from the dish.
A high quality LNB, with a noise level of $\sim$1.2\,dB,
is especially important for acceptable reception of digital signals. 
In order to receive both analogue and digital signals in both the 
``low'' and ``high'' parts of the Ku-band, a ``Universal LNB'' is required. 
A Universal LNB switches from low-band to high-band upon application of a 22\,kHz
signal from the receiver to the LNB and to horizontal and vertical 
polarisations upon application of 13\,V or 18\,V to the LNB's power supply.
All such signals utilise the co-axial readout cable from the LNB to the 
receiver such that no separate cable is required to power the LNB. 

A ``Twin LNB'' may be mounted on the dish.
The two LNB's may be slightly offset in angle to receive signals 
from two different satellites.
Alternatively they may be aligned with the same satellite thereby enabling 
two receivers to operate using the same dish, and hence to
receive digital and analogue signals independently with either 
horizontal or vertical polarisations.
Dish diameters quoted for single feed dishes should be increased by typically
20\% for dishes with dual feeds.
A ``Quatro LNB'' permits the simultaneous reception of low-band and high-band 
signals with both horizontal or vertical polarisations.
Quatro LNB's are typically used when the dish is part of a   
Satellite Master Antenna Television (SMATV) setup in which a single dish
supplies amplified signals to multiple receivers.  

\subsection{Viewing and Feedback} 

The signal from the dish passes through a 75\,$\Omega$ impedance coaxial 
cable to an analogue and/or digital receiver.
The cable and connectors should be of high (satellite) quality since the 
use of standard (terrestial TV) components can result in noise or signal loss.
If the cable length exceeds $\sim 30$\,m or if the dish is driving multiple devices, 
then additional amplification may be necessary.

The signal is decoded by a analogue and/or digital receiver and converted to 
a signal format (either PAL/SECAM or NTSC) which is compatible with the 
television set or Video Cassette Recorder.
Subtitles may prove useful for translations of the audio signal or even for 
transcription in the same language to facilitate understanding for viewers 
who read a given language well but are less exposed to it aurally. 
The most prevalent subtitle standard by far is the TeleText system developed 
by the BBC and IBA in the UK.
There is little widespread use of other systems such as the Closed Captioning system 
developed in the USA or the Antiope system developed in France.

For the reception of data rather than video signals, a data aquisition
system is required.
This could consist of a commodity PC with an additional, and readily 
available, commercial card for receiving the signal and decoding it.  
While the availability of video programs on desktop computers may be 
an advantage it is not a prerequisite for an STV service in HEP.
Incidentally, it is predicted that PC's and TV's will remain as distinct devices, 
at least for the next five years or so~\cite[page 13]{DTT98A}.   

Audio and fax feedback from viewers is possible using standard telephone lines.
This is supplemented by Internet-based tools such as E-mail, videoconferencing, 
whiteboards, etc. which provide more possibilities than mere audio feedback.
At dedicated centres each viewer may be provided with a simple numeric keypad 
with a built in microphone~\cite{DEC_MEYER}.
These are multiplexed together and connected by a single telephone 
line to the production centre.
This system permits all viewers to speak to the production centre,
and for the full audience to respond simultaneously to simple numeric questions 
or to arbitrary questions with multiple choice answers.
This enables the presenter to target the presentation to the specific needs 
of the audience, for example to respond to individual questions or to reiterate 
an argument if a significant fraction of the audience had trouble following it.

\subsection{Comparison with Internet\label{whynotinternet}}

Today a major transatlantic optical fibre has approximately ten times the capacity 
of the highest powered satellite and fibre systems already in the design phase 
will provide bandwidth equivalent to approximately 50 such satellites~\cite{ALEWINE_SHORT}.
Compared to fibre networks, satellite transmissions have higher bit error rates, 
a generally higher (but fixed) latency, and (usually) a higher cost. 
It is therefore reasonable to ask why one would ever consider using satellites 
in preference to the Internet.
For certain specific applications, however, the use of satellites is attractive
due to features which are complementary to fibre systems.
\begin{myitemize}
\item {\bf{Transmissions are broadcast to arbitrary numbers of recipients.}}  
      Satellite broadcasting offers the possibility of transmitting 
      video or other data of general interest to arbitrary numbers of 
      collaborating institutes worldwide, in a highly parallel fashion.      
      This is in contrast to the point-to-point nature of fibre-based networks.
      While the development of multicast systems on Internet 
      may ultimately provide similar functionality, this is not yet 
      available nor is the actual cost known.
\item {\bf{Transmissions are terrain-independent and cover remote regions.}}  
      The inadequacy of fibre links to remote regions is a 
      problem for many HEP institutes. 
      Even for regions with adequate national network performance, there are 
      frequently problems due to congestion on trans-oceanic 
      links~\cite{ICFA_NTF_REPORT_JULY98,WILLIAMS_CHEP98}. 
\item {\bf{Satellite bandwidth is guaranteed.}}  
      There is no competition for bandwidth on a dedicated satellite link. 
      Internet fibre links, however, often pass through a few tens of routers, 
      any one of which may become congested and start to drop a 
      significant fraction of the TCP/IP packets,
      with disasterous results for real-time signals such as video data. 
      This is a key issue for future networks such as Internet2 which aim 
      to provide guaranteed ``Quality of Service'' for critical applications.
      This is not yet realised, however, and the pricing structure of 
      such services is unclear.
\end{myitemize}
%

%
\section{Analysis of a HEP Satellite Television System\label{hepsat}}
%
%
In order to identify the critical issues, parameters, and cost involved in
the exploitation of STV in HEP, we analyse a specific deployment scenario
which we subsequently refer to as ``HEPSAT''.
While HEPSAT is not unreasonable it should certainly not be seen as a
fixed definition of a system which might ultimately be deployed.  

\subsection{Program content}

The most obvious use of an STV system in HEP is for broadcasting of video data.
We estimate very approximately the following potential broadcasting times:
\begin{myitemize}
\item 500\,hours/year for seminars                     (1\,hour/seminar    $\times$ 10\,seminars/week $\times$ 50\,weeks/year);
\item 240\,hours/year for major conferences            (3\,days/conference $\times$ 10\,conferences/year);
\item 640\,hours/year for major collaboration meetings (2\,days/meeting    $\times$ 4\,meetings/year$ \times$ 10\,collaborations);
\item 400\,hours/year for training                     (8\,hours/week      $\times$ 50\,weeks/year);
\end{myitemize}
These sum to a total time of approximately 1800\,hours/year or 20\% of the year.
Since no time of day is universally convenient due to the differing time zones 
of the viewers or conflicts in schedule, it is probably necessary to 
rebroadcast recordings of a significant fraction of the material, 
although for such programs real-time feedback is no longer possible.
Clearly our estimates can be adjusted somewhat depending on the 
perceived interest in the material and the amount of rebroadcasting.
Nonetheless, we estimate that reasonable bounds on video usage, 
including repeat broadcasts, are $1800 - 2700$\,hours/year or 
$20 - 30$\% of the year. 

In addition to video programs, it is of interest to examine potential 
broadcasting of certain types of experimental data over satellite.
The Babar, CDF, D0, RHIC and KLOE, NA48, NA49, Compass and other experiments 
will each acquire hundreds of Terabytes per year within the next 
few years while the ATLAS, CMS, and ALICE experiments at the LHC, 
starting in 2005, each plan to store more than 
1\,PB/year~\cite{ICFA_NTF_REPORT_JULY98}.
It is likely that a significant fraction of each experiments full data 
set will be transferred from the experiment to a few regional computing 
centres and that a smaller fraction will be transferred to the  
large number of collaborating institutes.
Much of the data is common to many sites, such as ``hot'' event samples, 
calibration constants, and even software updates. 
Given the point-to-multipoint nature of satellite transmissions it may prove
effective to broadcast such common data to all collaborating institutes 
rather than make multiple point-to-point transfers over the Internet.

The system could also be used for distributing HEP publicity material 
to the press but this would not need a significant fraction of the 
total bandwidth required~\cite{CALDER97A}.
While there are many interesting potential applications in the area of 
educational outreach, these are beyond the scope of this paper.

\subsection{Bandwidth Requirements}

The bandwidth required for good quality video and audio is 1.54\,Mbps, 
which is the smallest quantum generally available from satellite service providers. 
The HEP requirements for video are $\sim 25$\% of a full year while those 
for data are rather open-ended. 
The on-demand use of satellite bandwidth, rather than the use of a dedicated 
24\,hour/day service is, approximately three times more expensive 
per unit time~\cite{GLOBALACCESS_COSTS} such that on-demand usage for $\sim 33$\% of 
the time costs the same as a dedicated channel running 100\% of the time.
Moreover, on-demand usage requires considerably more advance coordination 
with the satellite service provider. 
Therefore, we assume that the system is available to HEP 
for 24 hours/day throughout the entire year.
This means that the bandwidth available for non-video applications is 
$\sim 75$\% of the total, which is sufficient to broadcast  
$\sim 5$\,Terabytes of data per year. 
Alternatively, one might envisage the transmission of non-HEP video programs
as part of a wider cost-sharing arrangement.

\subsection{Geographical coverage}
\begin{figure}[!htbp]
\begin{center}
    \mbox{\epsfig{file=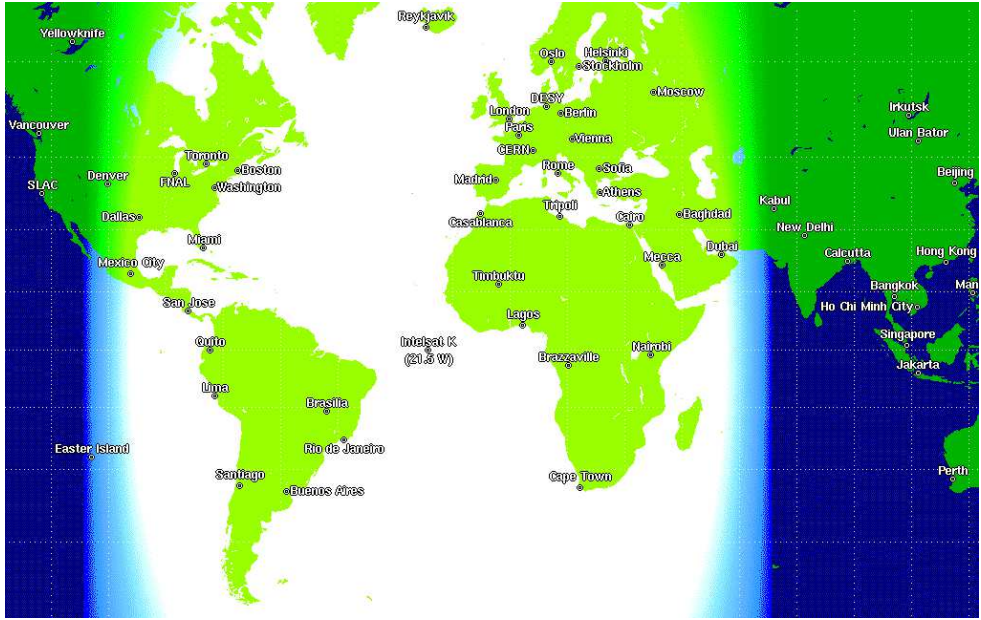,clip=}}
  \caption{Location of the Intelsat-K satellite showing the 
           coverage (lighter regions) for isotropic transmission.\label{fig:map}}
\end{center}
\end{figure}

The coverage considered includes Europe, part of the FSU, N. America, and S. America. 
We do not consider Asia, which is $\sim 10$\% of HEP, due to the
less homogeneous technical and regulatory issues 
(a system already running in Japan is described elsewhere~\cite{JAPAN_HEPSAT}).  
These requirements can be satisfied by a single satellite, 
such as Intelsat K at $21.5\degrees$W~\cite{GLOBALACCESS_COSTS}.
Figure~\ref{fig:map} shows the location of this satellite and
the coverage possible for isotropic transmission.
In practice, the satellite beams are focussed on the main areas of habitation
in Europe, North America, and in South America.
The peripheral loss of coverage shown in Figure~\ref{fig:map} is
nonetheless approximately valid.

The lack of coverage for California is unfortunate for HEP 
in general and SLAC in particular. 
It is exacerbated by the significant shielding effect of the Rocky mountains 
such that were one were to cover California by transmitting from a more 
westerly satellite, there would be a significant loss of easterly coverage.
Fortunately, California has excellent terrestrial fibre networks so
broadcasts to and from California could probably pass via the Internet to 
a more easterly location such as an uplink station at FNAL.

\subsection{Uplinks}
It is assumed that there are three uplinks, with two in Europe and one in the USA.  
Obvious candidates are CERN, DESY, and FNAL. 
Since almost all broadcasts would originate in one of several major 
HEP centres, there are no strong requirements on the mobility 
of the uplink equipment.
For non-standard sites, such as conferences which are not held at a major 
laboratory, the production equipment could be hired as a mobile unit, together 
with the satellite uplink equipment and qualified operators.  
Depending on local conditions and cost it may be advantageous to use a fibre-optic 
connection to a local ``uplink provider'' thereby obviating the need for 
dedicated uplink equipment.  
The HEP needs for production (studio) equipment are relatively modest, involving 
several video cameras, studio lights, and microphones at each site, 
in a fairly static configuration.

\subsection{Viewers}

In total, the HEP community comprises approximately $12\,000$ PhD. 
physicists (both experimental and theoretical), of which approximately
half are involved in research programs at CERN~\cite{NORDBERG_HEP_CENSUS}.
In addition there are $\sim$ 5\,000 HEP PhD. students~\cite{ECFA_HEP_CENSUS,NORDBERG_HEP_CENSUS}. 
It is assumed that there are $\sim$300 HEP institutes within the 
geographical region defined above~\cite{NORDBERG_HEP_CENSUS,ECFA_HEP_CENSUS,CANADA_HEP_CENSUS,DOVA_HEP_CENSUS}.
These estimates suffer from a lack of any worldwide survey, 
uncertainties in the definitions of ``physicist'' and ``HEP institute'',
and incomplete or out-of-date information.
However, for the purposes of this study only approximate numbers 
are required.
There should be (at least) one reception system in each HEP institute, 
made from readily available commodity components.
Feedback from viewers would be provided by a combination of  systems
based on the existing Internet and telecommunications infrastructure.

\subsection{Cost Estimate}

The coverage required for this study is Europe, the western part of the 
FSU, N. America, and S. America, which can be satisfied by a single satellite.
The cost of broadcasting over satellite to these regions at 1.54\,Mbps 
for 24\,hours/day throughout the year is 
$\sim 560$\,k\$/year~\cite{GLOBALACCESS_COSTS}
(there is modest drop in the unit cost for higher bandwidth requirements,
for example a 32.8\,Mbps broadcast costs $\sim 8\,000$\,k\$/year, {\em{i.e.}} 
$244$\,k\$/Mbps/year, compared to
$363$\,k\$/Mbps/year for the 1.54\,Mbps option~\cite{GLOBALACCESS_COSTS}).
It is important to note that this price corresponds to sufficient transmission 
power for acceptable reception on a commodity dish antenna of typically
less than 1\,m in diameter.

The cost of each fixed uplink station, including installation and commissioning, 
is $\sim 400$\,k\$~\cite{GLOBALACCESS_COSTS}.
Thus the cost for the three uplinks foreseen is $\sim 1\,200$\,k\$.
The personnel required is estimated to be 1 FTE for overall 
management and 1\,FTE per uplink site. 
The personnel costs are not included in this estimate.

The cost of each reception system is conservatively estimated to be 
$\sim 5$\,k\$/system~\cite{DEC_MEYER}.
This includes the purchase, transportation, installation, and commissioning
of the equipment required at the viewing site, in particular
the dish, LNB, cabelling, analogue/digital decoder, television set,  
video recorder, and a commercial card for receiving data on a PC.
Since the PC can be used for other applications most of the time
it is not included in the cost estimate. 
Institutes interested only in video programs do not need a PC at all.

Whereas viewers could purchase and install the equipment themselves,
the diffuse funding structure of HEP and the need for some central 
coordination would make such an approach less attractive than a centrally
coordinated and funded approach. 
The Internet and telecommunications systems used for viewer feedback are 
assumed to exist as part of the HEP infrastructure 
and are not included in the cost estimate.

\begin{figure}[!htbp]
\begin{center}
    \mbox{\epsfig{file=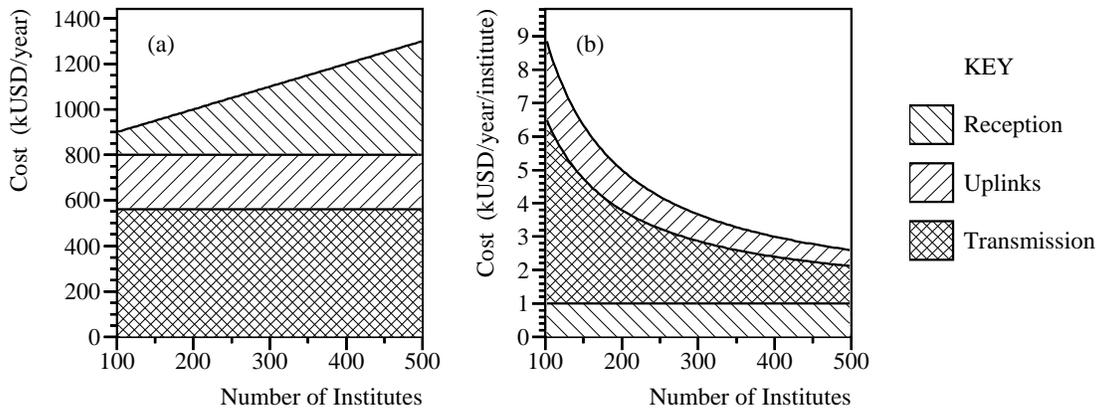,width=0.85\textwidth,clip=}}
  \caption{The total annual cost to HEP of the HEPSAT system 
           as a function of the number of participating institutes (a), 
           and the corresponding annual cost per institute (b).\label{fig:sat_costs}}
\end{center}
\end{figure}

In order to estimate the overall cost exposure we amortise the costs over five years. 
Figure~\ref{fig:sat_costs}(a) shows the total annual cost to HEP as a function of 
the number of participating institutes, while Figure~\ref{fig:sat_costs}(b) 
shows the annual cost per institute supposing that each institute contributes
to the total cost on a {\em{pro rata}} basis.
Assuming that 300 institutes participate, the total cost is 1.1\,M\$/year or 
an average of 3.7\,k\$/year/institute.
Approximately half of the cost is for the satellite bandwidth.
The remainder is required for HEP-specific hardware and is almost evenly 
split between the uplinks at the major centres 
and the reception systems in the institutes.
%

%
\section{Summary\label{summary}}
%
%
The potential application of satellite television systems in the 
global HEP community has been analysed.
There appears to be no technical or regulatory obstacle preventing 
the broadcast of video programs, such as seminars, conferences,
meetings or training sessions, as well as data to HEP institutes worldwide.
Operational experience within the Japanese HEP community
has demonstrated that such a service is useful to 
physicists~\cite{JAPAN_HEPSAT}. 

A cost estimate of the ``HEPSAT'' system, based on non-contractual 
quotes, determines that the total cost of deploying a 24\,hour/day 
dedicated system is $\sim$1.1\,M\$/year or $\sim$3.7\,k\$/year/institute, 
assuming that 300 institutes participate.
We believe this is not prohibitively expensive.
Quite apart from the intrinsic value of such a system to HEP, 
there are significant potential savings in time, travel costs, and 
conventional communications costs.

%
\section*{Acknowledgements}
%
%
We would like to thank Nick Meyer of DEC, 
Switzerland and Phil Govern and Margaret Baczor of Global Access, UK 
for invaluable discussions on many of the technical and financial aspects of 
satellite broadcasting.   

%
{\renewcommand{\baselinestretch}{0.4}

}


%
%

\small
\begin{thebibliography}{10}

\bibitem{ELBERT97A}
{B.R. Elbert}.
\newblock {\em {The Satellite Communications Applications Handbook}}.
\newblock Artech House, Boston, 1997.

\bibitem{PASCALL97A}
{S. Pascall and D.J. Withers}.
\newblock {\em {Commercial Satellite Communication}}.
\newblock Focal Press, Oxford, 1997.

\bibitem{BUNN_CHEP97}
{J. Bunn}.
\newblock {Collaborative Computing Environments for HEP}.
\newblock In {\em Invited talk at the CHEP'97 Conference}, Berlin, Germany,
  April 7-11 1997.
\newblock {\mbox{\tt{http://www.ifh.de/CHEP97/chep97.html}}}.

\bibitem{ICFA_NTF_REPORT_JULY98}
L.~Price et~al.
\newblock Status Report of the ICFA Networking Task Force, {\bf{ICFA/98/671}},
  July 1998. \\ (Available online:
  {\tt{http://nicewww.cern.ch/$\sim$davidw/icfa/july98report.html}}).

\bibitem{WILLIAMS_CHEP98}
{D.O. Williams, ``Networking needs and prospects: ICFA NTF report'' CHEP'98,
  Chicago, Aug 31 - Sept 4, 1998.}

\bibitem{TPN_MAY97_SHORT}
G. Kafka, in Telecom Product News/Satellite Communications, May/June 1997
  issue.

\bibitem{WWW_SATMAGAZINE}
SatMagazine, March 30, 1998, {{\tt{http://www.satmagazine.com/}}}.

\bibitem{ALEWINE_SHORT}
B.C. Alewine, keynote speech at Satellite 98,
  {{\tt{http://www.comsat.com/corp/home/bca\_speech.html}}}.

\bibitem{WWW_SATCODX}
{{\tt{http://www.satcodx.com/}}}.

\bibitem{WWW_ASTRA}
{{\tt{http://www.astra.lu/}}}.

\bibitem{WWW_EUTELSAT}
{{\tt{http://www.eutelsat.com/}}}.

\bibitem{WWW_INTELSAT}
{{\tt{http://www.intelsat.com/}}}.

\bibitem{WWW_HUGHES}
{{\tt{http://www.hughes.com/}}} and {{\tt{http://www.hughespace.com/}}}.

\bibitem{WWW_PANAMSAT}
{{\tt{http://www.panamsat.com/}}}.

\bibitem{WWW_LORAL}
{{\tt{http://www.loral.com/}}}.

\bibitem{WWW_ORION}
{{\tt{http://www.orionnetworks.net/}}}.

\bibitem{WWW_LOCKHEED}
{{\tt{http://www.lmco.com/}}}.

\bibitem{WWW_INTERSPUTNIK}
{{\tt{http://www.intersputnik.com/}}}.

\bibitem{WWW_COMSAT}
{{\tt{http://www.comsat.com/}}}.

\bibitem{GLOBALACCESS_COSTS}
{{\tt{http://www.globalaccess.com/}}. Private communication, P. Govern and M.
  Baczor, Global Access, UK}.

\bibitem{WWW_WILLIAMS}
{{\tt{http://www.twc.com/}}}.

\bibitem{WWW_SEPMEIER_97A}
W. Sepmeier, ``Worldwide TCP/IP Using Satellites'', SatMagazine, March 1998,
  {{\tt{http://www.satmagazine.com/}}}.

\bibitem{DTT98A}
{DTT Consulting, Winchester, UK, ``New Directions in Multimedia -- Internet via
  Satellite'', available from: {{\tt{http://www.spotbeam.com/}}", US\$ 995}}.

\bibitem{DEC_MEYER}
{Private communication, N. Meyer, BTV European Events Manager, DEC,
  Switzerland.}

\bibitem{CALDER97A}
{Private communication, N. Calder, Head of CERN Press Office}.

\bibitem{JAPAN_HEPSAT}
{T.Nakamura {\em{et al.}}, ``Broadcasting Physics seminars by using a
  satellite'' CHEP'98, Chicago, Aug 31 - Sept 4, 1998.}

\bibitem{NORDBERG_HEP_CENSUS}
{Private communication, M. Nordberg, CERN Office of Strategic Planning}.

\bibitem{ECFA_HEP_CENSUS}
{``The 1995 ECFA Survey of Particle Physics Activities and Resources in the
  CERN Member States'', ECFA/RC/96/244, 30 November 1996; ``The 1988 ECFA
  Survey of Particle Physics Activities and Resources in the CERN Member
  States'', ECFA/RC/90/178, 30 March 1990}.

\bibitem{CANADA_HEP_CENSUS}
Private communication, H. Mes, based on data from the I.P.P., Canada.

\bibitem{DOVA_HEP_CENSUS}
Private communication, M.T. Dova, Universidad Nacional de La Plata, Argentina.

\end{thebibliography}
%
\end{document}